%% file: AVERY.tex
\documentclass[acmtcps, nonacm]{acmart}
\renewcommand\footnotetextcopyrightpermission[1]{}

\usepackage[T1]{fontenc}
\usepackage{textcomp}
\usepackage{xcolor}
\usepackage{url}
\usepackage{balance}

\usepackage{amsmath,amssymb,amsfonts}
\usepackage{mathtools}

\usepackage{graphicx}
\usepackage{pgfplots}
\pgfplotsset{compat=newest}
\usetikzlibrary{patterns,shapes.arrows}
\usepackage{tikz}
\usepackage{svg}
\usepackage[font=footnotesize]{caption} 

\usepackage{booktabs}
\usepackage{multirow}
\usepackage{tabularx}
\usepackage{makecell}
\usepackage{adjustbox}
\usepackage{tablefootnote}

\usepackage{algorithm}
\usepackage{algpseudocode}
\algrenewcommand\algorithmiccomment[1]{\hfill\(\triangleright\)\;\parbox[t]{.6\linewidth}{#1}}
\makeatletter
\usepackage{comment}
\usepackage{academicons}
\usepackage{booktabs}
\usepackage{makecell}
\usepackage[table]{xcolor}
\usepackage{pifont}

\newcommand{\cmark}{\ding{51}}
\newcommand{\xmark}{\ding{55}}

\newcommand{\supc}[1]{%
    (\tikz[baseline=(char.base)]{
        \node[shape=circle, fill=black, text=white, inner sep=0.5pt] (char) {\small #1};
    })
}

\DeclareRobustCommand{\circlednum}[1]{%
    \tikz[baseline=(char.base)]{
        \node[
            shape=circle,
            fill=black,
            text=white,
            inner sep=1pt
        ] (char) {\fontsize{7pt}{7pt}\selectfont #1};
    }%
}

\makeatletter
\let\@authorsaddresses\@empty
\makeatother
\begin{document}

\newcommand{\papername}{AVERY}

\title{\papername{}: Intent-Driven \underline{A}daptive \underline{V}LM Split Computing via \underline{E}mbodied Self-Awareness for Efficient Disaster \underline{R}esponse S\underline{y}stems}



\author{Rajat Bhattacharjya}
\email{rajatb1@uci.edu}
\orcid{1234-5678-9012} 
\author{Sing-Yao Wu}
\email{email@uci.edu}
\author{Hyunwoo Oh}
\email{email@uci.edu}
\affiliation{%
  \institution{University of California, Irvine}
  \city{Irvine}
  \state{California}
  \country{USA}
}
\author{Chaewon Nam}
\authornote{Work done while at the University of California, Irvine.}
\author{Suyeon Koo}
\authornotemark[1]
\affiliation{%
  \institution{Kookmin University}
  \city{Seoul}
  \country{South Korea}
}
\author{Mohsen Imani}
\author{Elaheh Bozorgzadeh}
\author{Nikil Dutt}
\email{dutt@uci.edu}
\affiliation{%
  \institution{University of California, Irvine}
  \city{Irvine}
  \state{California}
  \country{USA}
}
\thanks{Paper is currently under review. Our code and dataset will be made public upon acceptance. Corresponding authors: Rajat Bhattacharjya (\texttt{rajatb1@uci.edu}) and Nikil Dutt (\texttt{dutt@uci.edu}). Authors' version posted for personal use and not for redistribution.}
\renewcommand{\shortauthors}{Bhattacharjya et al.}









\input{tex/01_abstract}


\maketitle

\input{tex/intro2}
\input{tex/back3}

\input{tex/prob2}
\input{tex/05_Experiments}
\input{tex/07_conclusion}
\input{tex/ack} 
\bibliographystyle{unsrt}

\input{tex/AVERY.bbl}



\end{document}

%% file: tex/01_abstract.tex
\begin{abstract}
    Unmanned Aerial Vehicles (UAVs) in disaster response require complex, queryable intelligence that onboard CNNs cannot provide. While Vision-Language Models (VLMs) offer this semantic reasoning, their high resource demands make on-device deployment infeasible, and naive cloud offloading fails under the low-bandwidth, unstable networks endemic to disaster zones. We present \texttt{AVERY}, an intent-driven adaptive split computing framework for efficient VLM deployment on resource-constrained platforms. \texttt{AVERY} is motivated by the observation that operator intent must be treated as a first-class system objective, since missions such as broad situational monitoring and precise, spatially grounded investigation require different semantic products, latency targets, and resource allocations. To reflect this, \texttt{AVERY} advances split computing beyond traditional depth-wise partitioning through a functional, cognitive-inspired dual-stream split: a high-frequency, low-resolution Context stream for real-time awareness, and a low-frequency, high-fidelity Insight stream for deep analysis. This design enables a hierarchical split strategy: computation is first separated by function, then partitioned depth-wise across edge and cloud when the Insight stream is required. A lightweight, self-aware onboard controller monitors network conditions and operator intent to select from pre-trained compression models, navigating the accuracy-throughput trade-off at runtime. Evaluated using LISA-7B in an edge-cloud setting under fluctuating network conditions, \texttt{AVERY} achieves 11.2\% higher accuracy than raw image compression, 93.98\% lower energy consumption than full-edge execution, and average accuracy within 0.75\% of the static High-Accuracy baseline during dynamic adaptation. Overall, \texttt{AVERY} enhances mission efficiency and enables real-time, queryable intelligence in dynamic disaster environments.
\end{abstract}

%% file: tex/intro2.tex
\vspace{-1mm}
\section{Introduction}
\label{sec:intro}
Disaster Response Systems (DRSs) are increasingly deploying Unmanned Aerial Vehicles (UAVs) to enhance situational awareness, support search-and-rescue operations, and assess infrastructure damage in large-scale emergencies such as floods, wildfires, and earthquakes~\cite{dis,tcps1,tro}. These UAVs act as autonomous agents~\cite{dronefl, sens, at} capable of rapidly traversing complex and dynamic environments, collecting high-resolution sensor data---often in the absence of reliable human supervision or stable communication infrastructure. Classical onboard intelligence typically relies on \textit{Convolutional Neural Networks} (CNNs)~\cite{cnndis} for basic scene understanding, offering computational efficiency under the strict energy and compute budgets of embedded hardware. However, such pipelines often produce large numbers of generic detections and labels, increasing operator cognitive load while failing to prioritize what matters most for the immediate mission.

\begin{figure*}[t] 
\centering 
\includegraphics[width=0.9\textwidth]{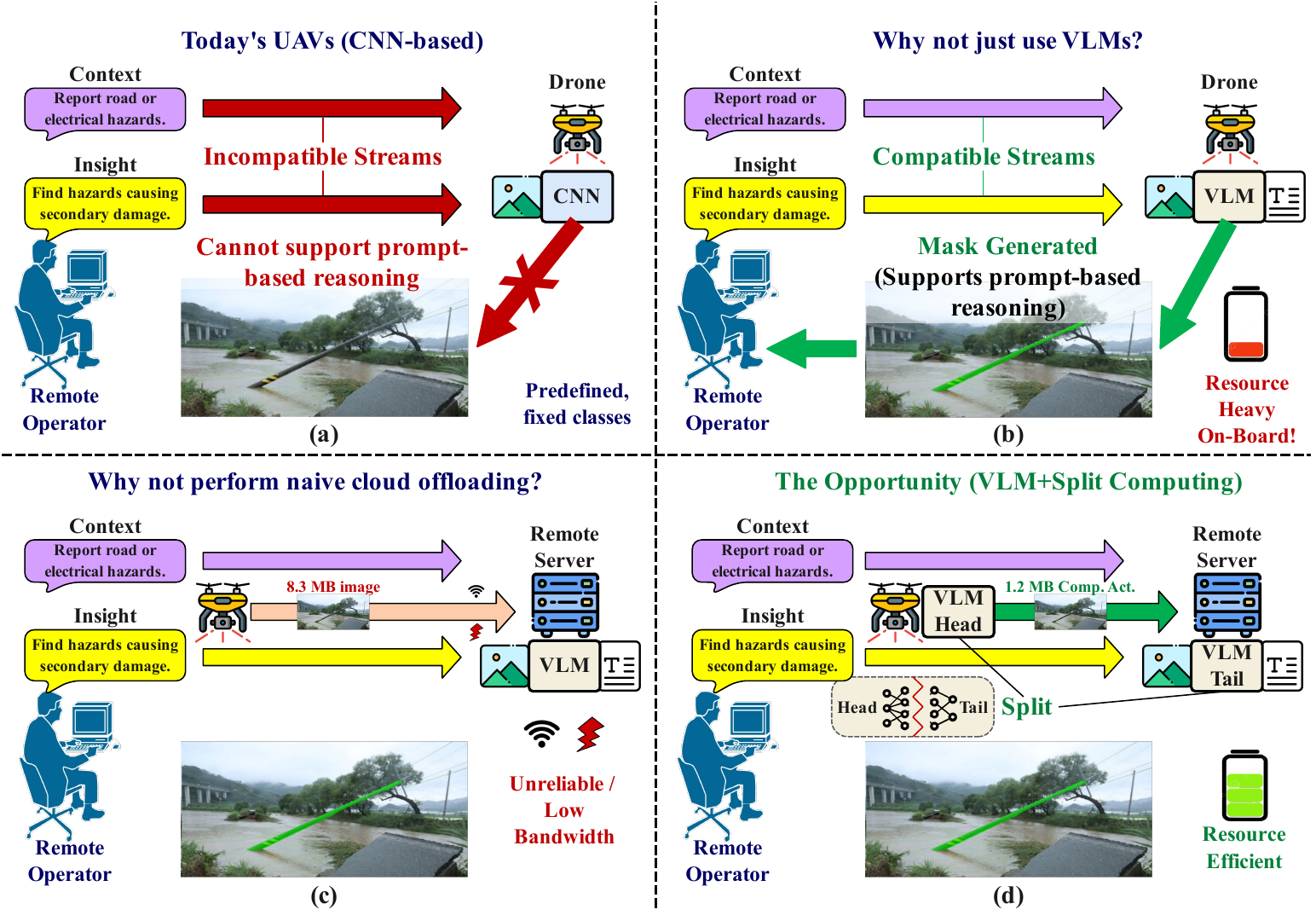} 
\vspace{-2mm}
\caption{The Motivation and \texttt{AVERY} Paradigm. (a) Conventional CNNs cannot support the distinct, prompt-based reasoning needed for multi-level disaster response, such as broad \texttt{Context} queries for triage and specific \texttt{Insight} queries for investigation. (b) While a full on-device VLM can process these queries, its prohibitive energy cost makes it infeasible for UAV deployment. (c) Naive cloud offloading fails under the unreliable, low-bandwidth networks found in disaster zones. (d) \texttt{AVERY} resolves these conflicts with a dual-stream (context and insight) split computing architecture, intelligently transmitting either lightweight context features or compressed insight activations to enable hierarchical, efficient, multi-level intelligence.}
\vspace{2mm}
\label{fig:motivation} 
\vspace{-3ex}
\end{figure*}

Also, as illustrated in Figure~\ref{fig:motivation}(a), CNNs lack the \textit{semantic flexibility} and \textit{contextual reasoning} required to support operator-driven queries in complex disaster environments. For instance:

\begin{itemize} 
\item In \textbf{flood response}, a CNN may segment water regions but cannot answer a prompt like \texttt{``locate vehicles partially submerged but still accessible by road.''}

\item After an \textbf{earthquake}, detecting ``collapsed structures'' is relatively straightforward; determining \texttt{``where individuals are trapped near collapsed structures''} requires richer grounding of objects, relationships, and spatial cues.

\item In a \textbf{wildfire scenario}, a CNN might outline fire boundaries but cannot infer or highlight the \textit{safest reinforcement route} to contain spread.
\end{itemize}

These limitations point to a broader gap: disaster response requires not just detection, but \textit{actionable intelligence}. In practice, operators can express mission needs and intent through natural language (NL), and these intents can differ substantially in the semantic capability they demand. For example, a triage-oriented query such as \texttt{``What is happening in this sector?''} requires fast, coarse, high-refresh awareness, whereas a grounded query such as \texttt{``highlight individuals near submerged vehicles''} requires segmentation-aware, high-fidelity spatial reasoning. Thus, operator intent must be treated as a \textit{first-class system objective}, because different intents require fundamentally different semantic products rather than merely different points along a latency--accuracy trade-off.

\textit{Vision-Language Models} (VLMs)~\cite{vlmsurvey} offer a compelling alternative, enabling \textit{text-conditioned visual understanding}, \textit{open-vocabulary recognition}, and promptable reasoning---precisely the kind of flexible perception needed for next-generation DRSs. However, VLMs incur substantial compute and energy costs~\cite{energy}, making full on-device deployment impractical on UAV platforms (Figure~\ref{fig:motivation}(b)). Naive cloud offloading is also ineffective, since disaster environments often exhibit unreliable and bandwidth-limited connectivity (Figure~\ref{fig:motivation}(c)), leading to unstable throughput and prohibitive transmission latencies for high-resolution visual data.
This raises a fundamental question:

\textit{How can we enable intent-driven, semantically rich, bandwidth-aware, and energy-efficient VLM inference on UAVs operating in dynamic disaster environments?}


\textbf{Split computing, i.e., partitioning a model into a head (edge) and tail (cloud),} is a promising direction~\cite{splitz,matsubara2022bottlefit,part,spltg,parts}, but prior work has focused largely on depth-wise partitioning of CNN-style models with a single fixed inference pathway and output objective. Such approaches are not directly suited to VLM-powered disaster response, where operator queries can demand fundamentally different semantic products ranging from lightweight situational awareness to pixel-level grounded analysis. In this setting, the appropriate transmitted representation is determined not only by resource conditions, but first by the semantic function required by the task. We therefore argue that the split should not be determined by depth alone, but first by \textbf{function}. Specifically, we introduce a \textbf{dual-stream split} (Figure~\ref{fig:motivation}(d)) that separates the VLM into a high-frequency, low-resolution \textbf{Context Stream} for rapid situational awareness and a low-frequency, high-fidelity \textbf{Insight Stream} for deeper, segmentation-driven analysis.

\begin{figure*}[t]
\centering
\includegraphics[width=0.54\textwidth]{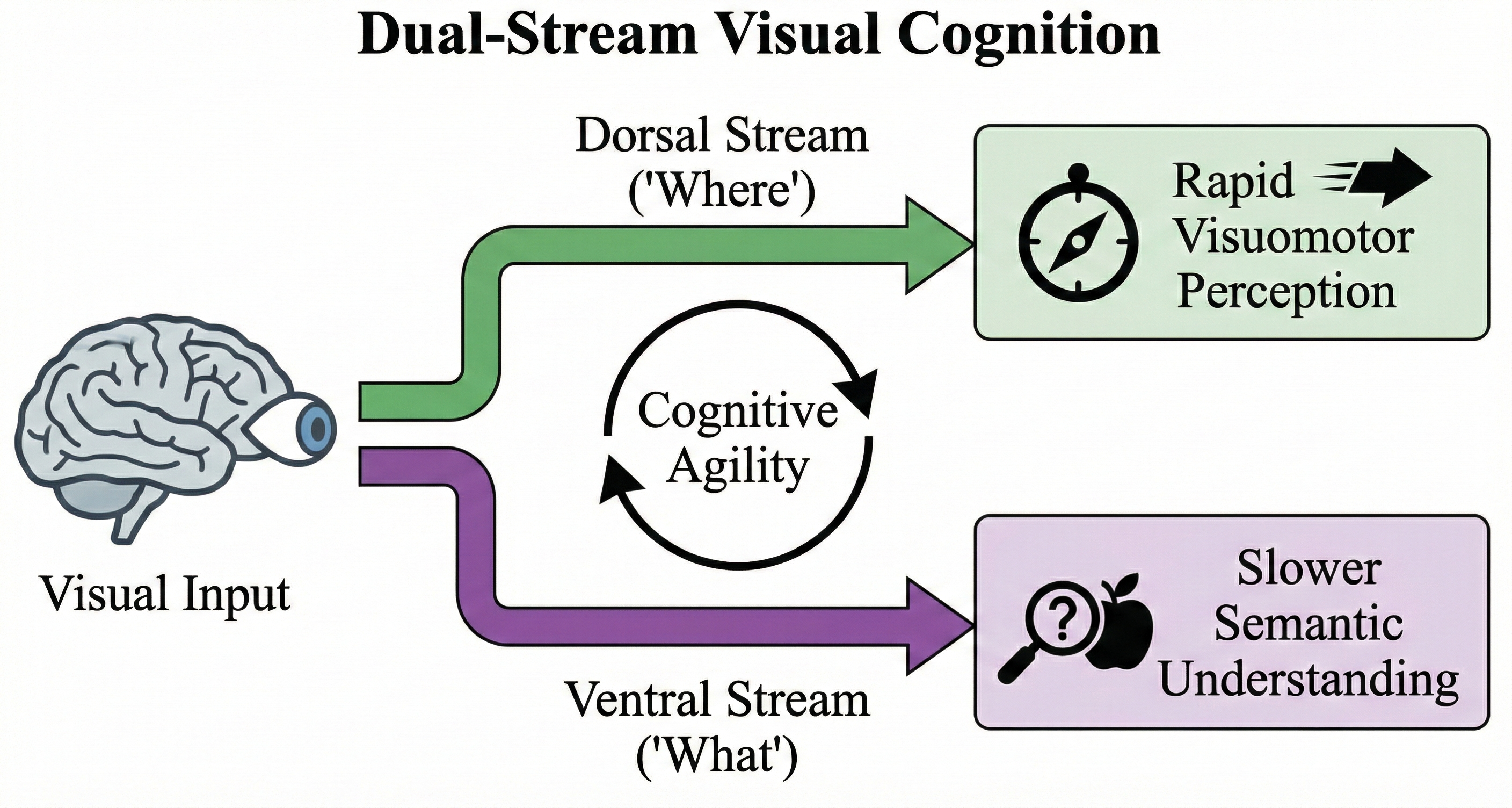}
\caption{{Dual Stream Organization for Human Visual Cognition. Dorsal Stream (in green): Fast Awareness, Ventral Stream (in purple): Slower Semantic Understanding.}}
\label{fig:dual}
\end{figure*}

This functional separation is inspired by the \textbf{dual-stream organization of human visual cognition}~\cite{neuro1,neuro2}, where rapid visuomotor awareness and slower semantic understanding proceed through complementary pathways as shown in Figure~\ref{fig:dual}. Mirroring this principle, the \textit{Context Stream} supports fast awareness, while the \textit{Insight Stream} supports grounded semantic reasoning. Crucially, stream selection in our system is not merely a performance optimization; it is an intent-conditioned capability decision. Operator (human) intent determines which stream is semantically admissible, while resource conditions such as bandwidth and onboard energy determine how that stream should be delivered.

Building on this, we present \texttt{AVERY}, which, to the best of our knowledge, is the first \textbf{intent-driven adaptive split computing framework for VLM-based disaster-response UAVs.} \texttt{AVERY} introduces a two-level hierarchical adaptive strategy: it first performs \textit{stream-level gating} based on operator intent, and then performs \textit{resource-level adaptation} within the selected stream through model-depth splitting and learned compression. In particular, when the computationally intensive Insight Stream is invoked, \texttt{AVERY} applies early edge--cloud partitioning together with learned bottleneck~\cite{matsubara2022bottlefit} compression to navigate the trade-off between semantic fidelity, throughput, and onboard energy consumption under dynamic network conditions. 

Formally, let $I_t$ denote operator intent at time $t$. Each intent induces semantic requirements and associated service-level objectives (SLOs), such as minimum fidelity and freshness constraints. Stream selection is therefore constrained by intent:
\[
\text{stream}_t \in \mathcal{S}(I_t),
\]
where $\mathcal{S}(I_t)$ denotes the set of streams capable of satisfying the semantic requirements of $I_t$. Resource adaptation then optimizes compression ratio and transmission behavior within this feasible set. In this way, \texttt{AVERY} elevates human mission requirements from a soft preference to a first-class systems constraint.

Our contributions are as follows:

\begin{enumerate}

\item \textbf{An Intent-Driven Adaptive Split Computing Framework for VLMs:} We present \texttt{AVERY}, the first framework to enable VLM deployment on resource-constrained UAVs for disaster response through intent-driven adaptive split computing. We show that, for VLM-based disaster intelligence, the key challenge is not only where to split by depth, but how to first organize computation by semantic function and then adapt that computation to dynamic resource conditions.

\item \textbf{A Practical Dual-Stream Operational Model for Disaster Response:} We introduce a dual-stream architecture that separates VLM inference into a high-frequency, low-resolution \textbf{Context Stream} for operator awareness and drone piloting, and a low-frequency, high-fidelity \textbf{Insight Stream} for the core segmentation-driven VLM analysis. This makes the system operationally aligned with the heterogeneous intent structure of real disaster-response workflows.

\item \textbf{A Lightweight, Self-Aware Controller for Intent-Conditioned Adaptation:} We design a lightweight, onboard controller that first selects the semantically admissible stream based on operator intent and then manages the resource-intensive Insight Stream through adaptive compression and model-depth partitioning. By monitoring real-time network conditions and high-level operator intent, the controller navigates a fundamental co-design trade-off among semantic fidelity, throughput, and on-device energy consumption.

\item \textbf{A Domain-Specific Dataset and Fine-Tuned VLM for Flood Response:} To enable realistic evaluation, we construct and release a new disaster-focused dataset, \texttt{Flood-ReasonSeg}, readily usable for VLM-based tasks. We use it to fine-tune the LISA VLM~\cite{lisa}, creating the first open-vocabulary, queryable segmentation model specifically tailored for identifying critical targets such as stranded individuals and vehicles in flood scenarios.

\end{enumerate}

In our evaluation on an NVIDIA Jetson AGX Xavier (32 GB) with a representative cloud backend under variable network conditions, \texttt{AVERY} consistently outperforms baselines, achieving \textbf{11.2\% higher accuracy} on average than raw image compression, \textbf{93.98\% lower energy consumption} compared to full onboard execution of the Insight Stream's segmentation backbone, and average accuracy within \textbf{0.75\% of the static High-Accuracy baseline} during dynamic adaptation. These results show that intent-driven stream selection combined with adaptive split computing enables resilient, semantically rich intelligence for mission-critical DRSs.

The remainder of this paper is organized as follows. Section~\ref{sec:background} reviews the technical background and positions \texttt{AVERY} relative to prior work. Section~\ref{sec:sysmodel} presents the system model for hierarchical runtime adaptation. Section~\ref{sec:architecture} describes the \texttt{AVERY} architecture and its adaptive control logic. Section~\ref{sec:experiments} evaluates the system through both static characterization and dynamic runtime experiments. Section~\ref{sec:conclusion} concludes the paper and outlines directions for future work.

%% file: tex/back3.tex
\section{Background and Related Work}
\label{sec:background}

\texttt{AVERY} builds on four technical threads: the semantic expressiveness of Vision-Language Models (VLMs), the efficiency of split computing, emerging edge--cloud execution frameworks for multi-modal and large language models, and the adaptability of embodied self-awareness. In mission-critical UAV deployments, however, utility is inherently intent-conditioned. Different operator queries require different semantic products, ranging from coarse situational awareness to pixel-accurate grounding, and therefore impose different requirements on timeliness, fidelity, and communication overhead. This motivates \texttt{AVERY}'s dual-stream design, where the Context and Insight streams provide semantically distinct execution modes rather than simply representing different operating points along a single timeliness--accuracy curve.

\subsection{Vision-Language Models for Grounded Disaster Perception}
\label{sec:background_vlm}

VLMs combine visual encoders with Large Language Models (LLMs), enabling prompt-driven reasoning over visual scenes and open-vocabulary understanding beyond fixed label sets~\cite{vlmsurvey,lisa}. This capability is particularly valuable in disaster response, where UAVs operate in unstructured and rapidly changing bandwidth-limited environments and must answer task-specific operator queries such as \texttt{``are there people near the submerged car?''} or \texttt{``highlight the stranded vehicle''}. Unlike conventional CNN-based pipelines, which are typically trained for fixed closed-set perception tasks, VLMs support semantically rich interaction and can produce heterogeneous outputs conditioned on natural-language intent. Similar bandwidth-limited field conditions also arise in scenarios such as military UAV reconnaissance~\cite{droneedge} and agricultural monitoring systems~\cite{farmbeats,shekhar}, further motivating efficient deployment of such models beyond disaster settings.

An important instance of this capability is grounded semantic perception, where a model must not only reason over scene content but also localize relevant entities spatially. LISA~\cite{lisa}, for example, augments a vision-language model with segmentation capability through a special \texttt{<SEG>} token, enabling promptable grounded mask generation. 
For disaster-response UAVs, such grounded outputs are critical when coarse awareness is insufficient and the operator must distinguish exact victims, hazards, or affected infrastructure. These works therefore motivate the \emph{semantic need} addressed by \texttt{AVERY}: a deployed UAV system must support both lightweight semantic triage and precise spatial grounding, depending on mission intent. At the same time, VLMs' substantial computational and energy demands~\cite{energy} make direct end-to-end deployment on resource-constrained UAVs impractical, motivating adaptive edge--cloud execution.

\subsection{Split Computing with Learned Compression}
\label{sec:background_split}

Split computing addresses the gap between large models and constrained edge devices by partitioning inference across an on-device ``head'' and a remote " tail " ~\cite {neurosurgeon,splitz}. Rather than transmitting raw sensor data, the edge executes early layers locally and sends intermediate activations to the server, reducing communication overhead and potentially improving privacy and latency. A large body of prior work studies depth-wise partitioning of CNN-style perception models and explores how the split point should be selected under changing resource conditions~\cite{neurosurgeon,splitz}.

A complementary line of work introduces learned feature compression, in which trainable bottlenecks are inserted around the split to reduce the size of transmitted activations while preserving task utility~\cite{bottlenetpp,matsubara2022bottlefit,ladon}. These results are highly relevant to \texttt{AVERY} and directly motivate our use of learned bottleneck compression within the Insight stream, which will be discussed later. However, existing split computing frameworks generally assume a fixed task semantic and optimize resource trade-offs within a single inference pathway. In contrast, \texttt{AVERY}'s first runtime decision is not merely \emph{where} to split, but \emph{which semantic pathway is admissible} for the current operator intent. The resulting control problem is therefore not purely resource-aware, but also conditioned on the semantic requirements of the current task.
\subsection{Edge--Cloud / Distributed Inference for VLMs, MLLMs, and LLMs}
\label{sec:background_distributed}

Recent work has begun to explore edge--cloud execution for VLMs, multi-modal LLMs (MLLMs), and LLMs more directly. Distributed VLMs place visual processing on edge devices and language generation on the server, showing that cloud--edge decomposition can improve throughput relative to cloud-only execution even without explicit model compression~\cite{distributed_vlms}. 
S2M3~\cite{s2m3} further investigates functional-level splitting and sharing for multi-modal models across tasks, illustrating that module-level decomposition can reduce memory footprint and latency on resource-constrained devices. These systems demonstrate that multi-modal inference can itself be distributed across heterogeneous compute resources.

Other systems push this direction further through adaptive or task-aware collaboration. SpotVLM~\cite{spotvlm}, for example, uses cloud--edge collaboration with context transfer to support real-time VLM inference under changing conditions. TMO~\cite{tmo} studies local--cloud inference offloading for multi-modal, multi-task, and multi-dialogue settings, emphasizing scheduling decisions across heterogeneous request types. MoA-Off~\cite{moaoff} similarly explores modality-aware adaptive offloading for MLLMs, showing that different modalities may impose different system bottlenecks and therefore benefit from differentiated execution strategies. Complementary work on collaborative edge-to-server VLM inference uses selective retransmission of task-relevant image regions to reduce communication cost while preserving inference quality~\cite{vlm_collab_edge_server}. Taken together, these works show growing interest in making large multi-modal models practical through distributed inference, adaptive offloading, and communication-aware collaboration.

These works are important and should be viewed as adjacent rather than orthogonal to \texttt{AVERY}. However, they typically treat execution as a monolithic or scheduler-level deployment problem: the model is offloaded, partitioned, or routed under resource constraints, but the semantic execution path itself is not organized around distinct operator intents. In other words, they largely optimize \emph{where} or \emph{how much} computation should be executed across edge and cloud, but do not explicitly structure execution around different semantic requirements of the task itself. \texttt{AVERY} instead introduces a functional dual-stream decomposition in which one pathway is optimized for timely semantic awareness and the other for grounded spatial analysis. Operator intent first determines which pathway is admissible; resource-aware adaptation is then applied within that pathway.

\subsection{Embodied Self-Awareness for Adaptive Control}
\label{sec:background_selfaware}

In disaster response, both system resources and environmental conditions can change rapidly during operation. UAVs may encounter fluctuating bandwidth, varying onboard compute-power budgets, and shifting mission demands, all of which directly affect what forms of perception and reasoning can be sustained at runtime. Embodied self-awareness refers to a system's ability to monitor such internal and external context and adapt its behavior accordingly to maintain effective operation~\cite{embodied}.

This adaptive capability is particularly important for split VLM execution. Static partitioning strategies cannot account for changing wireless conditions, onboard resource envelopes, or the differing semantic requirements induced by heterogeneous operator queries. In practice, some queries require only lightweight semantic awareness, while others require high-fidelity spatial grounding with stricter timeliness and fidelity demands. In \texttt{AVERY}, self-awareness therefore extends beyond resource sensing alone: the system incorporates operator intent as part of the runtime control context, since intent determines the semantic capability required from the model and the corresponding service targets. The controller must therefore jointly adapt stream selection, split execution, and compression policy based on bandwidth, onboard resource conditions, and intent-conditioned semantic demands.

\begin{table*}[t]
\centering
\caption{Positioning of \texttt{AVERY} relative to representative prior systems. We emphasize only the properties most relevant to \texttt{AVERY}'s novelty: \emph{runtime-adaptive execution}, \emph{semantically distinct execution modes}, \emph{functional dual-stream decomposition}, \emph{intent-gated pathway selection}, \emph{spatially grounded output}, and \emph{embodied disaster-response UAV deployment}. Here, \cmark\ denotes that the property is an explicit design feature of the system, while \xmark\ denotes that it is not a primary feature. In particular, "spatially grounded output'' refers to user-facing grounded outputs such as masks or precise localized regions, rather than internal region-of-interest (ROI) or attention-based mechanisms used only to assist inference.}
\label{tab:rw_compare}
\small
\setlength{\tabcolsep}{4pt}
\renewcommand{\arraystretch}{1.15}
\resizebox{\linewidth}{!}{
\begin{tabular}{l c c c c c c c}
\toprule
\textbf{Work} & \textbf{Family} &
\makecell{\textbf{Runtime}\\\textbf{Adaptive}} &
\makecell{\textbf{Distinct}\\\textbf{Semantic}\\\textbf{Modes}} &
\makecell{\textbf{Functional}\\\textbf{Dual-Stream}} &
\makecell{\textbf{Intent-Gated}\\\textbf{Pathway}} &
\makecell{\textbf{Spatially}\\\textbf{Grounded}\\\textbf{Output}} &
\makecell{\textbf{Disaster / UAV}\\\textbf{Embodied Setting}} \\
\midrule

\rowcolor{gray!10}
\multicolumn{8}{l}{\textbf{Grounded VLM perception}} \\
LISA~\cite{lisa} & VLM & \xmark & \xmark & \xmark & \xmark & \cmark & \xmark \\
\addlinespace[2pt]

\rowcolor{gray!10}
\multicolumn{8}{l}{\textbf{Split computing / learned compression}} \\
Neurosurgeon~\cite{neurosurgeon} & DNN/CNN & \cmark & \xmark & \xmark & \xmark & \xmark & \xmark \\
BottleNet++~\cite{bottlenetpp} & DNN/CNN & \xmark & \xmark & \xmark & \xmark & \xmark & \xmark \\
Ladon~\cite{ladon} & DNN/CNN & \xmark & \xmark & \xmark & \xmark & \xmark & \xmark \\
\addlinespace[2pt]

\rowcolor{gray!10}
\multicolumn{8}{l}{\textbf{Distributed multi-modal / LLM inference}} \\
Distributed VLMs~\cite{distributed_vlms} & VLM & \xmark & \xmark & \xmark & \xmark & \xmark & \xmark \\
S2M3~\cite{s2m3} & MLLM & \xmark & \xmark & \xmark & \xmark & \xmark & \xmark \\
SpotVLM~\cite{spotvlm} & VLM & \cmark & \xmark & \xmark & \xmark & \xmark & \xmark \\
TMO~\cite{tmo} & LLM/MLLM & \cmark & \xmark & \xmark & \xmark & \xmark & \xmark \\
MoA-Off~\cite{moaoff} & MLLM & \cmark & \xmark & \xmark & \xmark & \xmark & \xmark \\
\makecell[l]{Collaborative Edge-to-Server\\VLM Inference~\cite{vlm_collab_edge_server}} & VLM & \cmark & \xmark & \xmark & \xmark & \xmark & \xmark \\
\addlinespace[2pt]

\rowcolor{blue!8}
\textbf{\texttt{AVERY} (this work)} & \textbf{VLM} & \cmark & \cmark & \cmark & \cmark & \cmark & \cmark \\
\bottomrule
\end{tabular}
}
\end{table*}
\subsection{Positioning of \texttt{AVERY}}
\label{sec:background_positioning}

Taken together, prior work establishes the importance of grounded VLM perception, split inference under resource constraints, distributed execution of multi-modal models, and runtime-adaptive CPS behavior. However, these works typically address only subsets of the design space. Some support grounded semantic perception without edge--cloud adaptation; others support distributed or adaptive execution without semantically distinct execution modes or intent-conditioned pathway selection.

\texttt{AVERY} is designed at the intersection of these threads. In particular, it jointly combines: \emph{(i)} runtime-adaptive edge--cloud execution, \emph{(ii)} semantically distinct execution modes for timely awareness versus grounded analysis, \emph{(iii)} a functional dual-stream decomposition, \emph{(iv)} operator-intent-conditioned pathway selection, \emph{(v)} grounded output support, and \emph{(vi)} deployment in an embodied disaster-response UAV setting. Table~\ref{tab:rw_compare} summarizes this positioning relative to representative prior systems.

%% file: tex/prob2.tex
\vspace{-2mm}
\section{System Model for Hierarchical Runtime Adaptation}
\label{sec:sysmodel}

\begin{figure*}[t]
\centering
\includegraphics[width=0.95\textwidth]{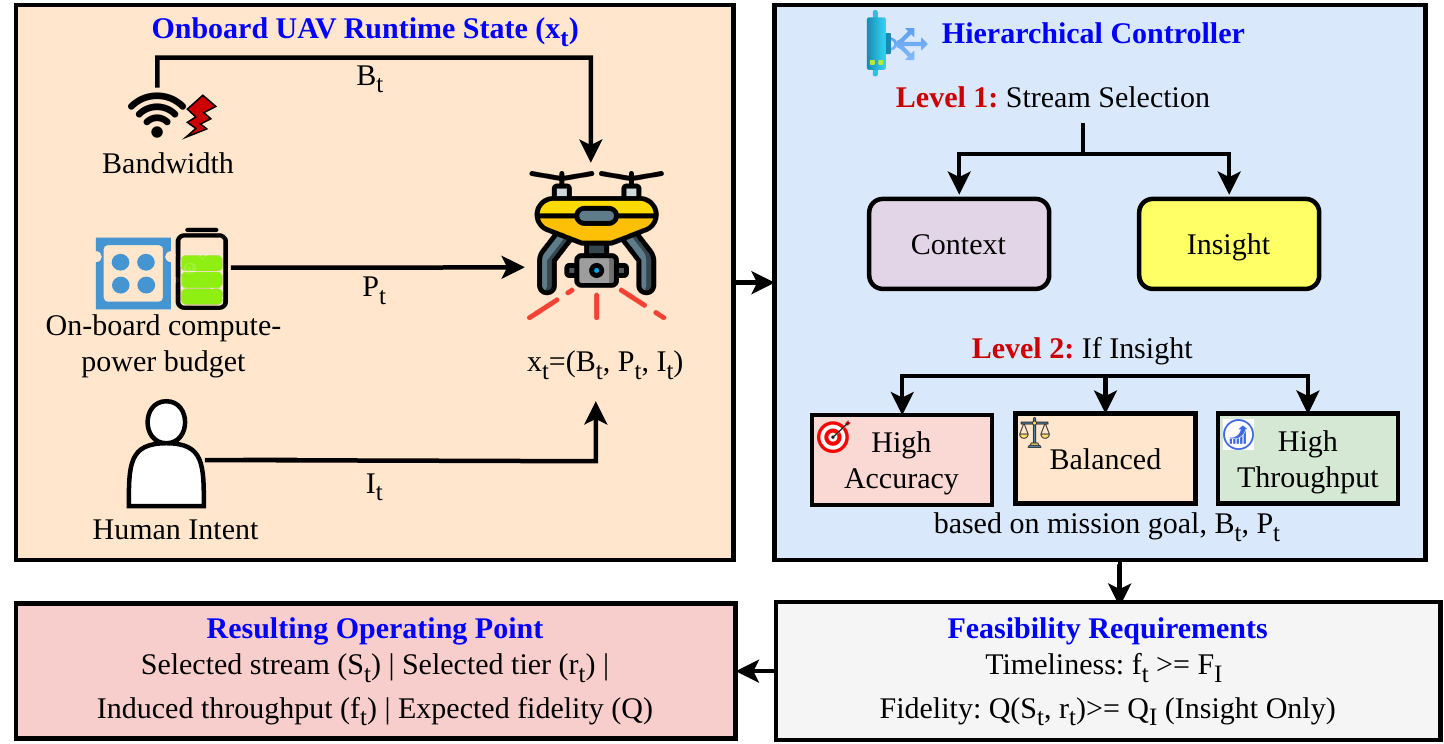}
\vspace{-2mm}
\caption{High-level system model of \texttt{AVERY}. The onboard controller observes the UAV runtime state $x_t=(B_t,P_t,I_t)$, where available bandwidth, onboard compute-power budget, and operator intent jointly define the current operating condition. The controller first selects the admissible stream and then, for Insight operation, selects a feasible operating tier based on mission goal and runtime resource conditions. The resulting configuration induces an update throughput $f_t$ and expected fidelity $Q$, which must satisfy the timeliness and fidelity requirements of the current task.}
\vspace{2mm}
\vspace{-4.5ex}
\label{fig:sysmod}
\end{figure*}

We model \texttt{AVERY} as a hierarchical adaptive split computing system operating under dynamic communication and onboard resource constraints, as shown in Figure~\ref{fig:sysmod}. At each decision epoch $t$, the UAV observes the runtime state and receives an operator prompt whose intent determines the semantic capability required by the mission. In \texttt{AVERY}, adaptation is hierarchical: operator (human) intent first selects the admissible semantic pathway, and resource-aware control then selects the operating point within that pathway. Table~\ref{tab:notation} summarizes the notation used throughout this section.

\subsection{UAV Runtime State and Operator Intent}

At each epoch $t$, the onboard controller observes the UAV runtime state
\[
x_t = (B_t, P_t, I_t),
\]
where $B_t$ denotes available uplink bandwidth, $P_t$ denotes the onboard compute-power budget, and $I_t$ denotes operator intent.

We distinguish two intent types, corresponding directly to \texttt{AVERY}'s dual-stream design:
\begin{itemize}
    \item \textbf{Context-level intents}: queries requiring coarse semantic understanding or situational triage, where a text-level response is sufficient.
    \item \textbf{Insight-level intents}: queries requiring fine-grained spatial grounding and precise masks.
\end{itemize}

Each intent $I_t$ induces service requirements:
\begin{itemize}
    \item a minimum update-timeliness target $F_I$, specifying the required rate at which updated information must reach the operator,
    \item for Insight-level intents, a minimum fidelity target $Q_I$, specifying the required quality of spatial grounding in the returned output.
\end{itemize}

In our setting, lower bandwidth increases transmission delay and therefore reduces how quickly updated scene information reaches the operator. We therefore operationalize update timeliness through achieved update throughput. Specifically, if a selected configuration delivers updates at rate $f_t$ (packets/s), then $f_t$ serves as a runtime proxy for update timeliness. Accordingly, $F_I$ denotes the minimum update-rate requirement associated with intent $I$. This requirement applies to both Context-level and Insight-level intents, although in practice it is most constraining for Insight-level intents, where the heavier Insight Stream introduces a stronger timeliness--fidelity trade-off. For Insight-level intents in our deployment, we instantiate this requirement as a minimum Insight update rate of 0.5 packets per second (PPS). This threshold is deployment-dependent and can be reconfigured for other settings.
\begin{table}[t]
\centering
\caption{Notation used in the system model}
\label{tab:notation}
\small
\begin{tabular}{l p{5.5cm}}
\toprule
\textbf{Symbol} & \textbf{Meaning} \\
\midrule
$t$   & Decision epoch at which the controller updates its decision. \\
$B_t$ & Available uplink bandwidth at epoch $t$. \\
$P_t$ & Onboard compute-power budget at epoch $t$. \\
$I_t$ & Operator intent at epoch $t$. \\
$x_t$ & UAV runtime state tuple, $x_t = (B_t, P_t, I_t)$. \\
$S_t$ & Selected stream, where $S_t \in \{\texttt{Context}, \texttt{Insight}\}$. \\
$\mathcal{S}(I_t)$ & Set of streams admissible for intent $I_t$. \\
$r_t$ & Selected operating tier for the Insight Stream. \\
$f_t$ & Achieved update throughput (packets/s), used as a runtime proxy for update timeliness. \\
$F_I$ & Minimum update-timeliness requirement for intent $I$. \\
$Q_I$ & Minimum fidelity requirement for Insight-level intent $I$. \\
$Q(S_t,r_t)$ & Expected output fidelity of the selected configuration. \\
$f(B_t,r_t,P_t)$ & Achievable throughput under bandwidth $B_t$, tier $r_t$, and onboard budget $P_t$. \\
\bottomrule
\end{tabular}
\end{table}

\subsection{Hierarchical Decision Model}

At runtime, the onboard controller selects:
\begin{itemize}
    \item stream $S_t \in \{\texttt{Context}, \texttt{Insight}\}$,
    \item for the Insight Stream, an operating tier $r_t$ chosen from a discrete set of pre-profiled configurations.
\end{itemize}

The first-level decision is functionally constrained by operator intent:
\[
S_t \in \mathcal{S}(I_t),
\]
where
\[
\mathcal{S}(I_t)=
\begin{cases}
\{\texttt{Context}\}, & \text{Context-level intent},\\
\{\texttt{Insight}\}, & \text{Insight-level intent}.
\end{cases}
\]

Thus, Context-level intents are served through the lightweight Context Stream, while Insight-level intents require the richer Insight Stream. The Context Stream is designed to satisfy the timeliness needs of Context-level intents through frequent lightweight updates, whereas the Insight Stream must explicitly trade update timeliness against fidelity due to its larger payload. Once the stream is determined, the controller selects the Insight operating tier based on mission goal, bandwidth, and onboard resource budget.

\subsection{Feasibility and Control Policy}

At each epoch $t$, the controller selects $(S_t,r_t)$, which induces an achievable update throughput
\[
f_t = f(B_t,r_t,P_t),
\]
such that
\[
S_t \in \mathcal{S}(I_t), \qquad
f_t \ge F_I,
\]
and, for Insight-level intents,
\[
Q(S_t,r_t) \ge Q_I.
\]

Here, $f(B_t,r_t,P_t)$ denotes the achievable update throughput under the current bandwidth, selected tier, and onboard compute-power budget, while $Q(S_t,r_t)$ denotes the expected output fidelity of that configuration. Higher-fidelity Insight tiers may deliver updates at lower rates, but remain admissible as long as they satisfy the minimum timeliness requirement of the current mission.

Importantly, the update-timeliness requirement is distinct from the bandwidth threshold of any particular operating tier. For Insight-level intents in our deployment, we require a minimum Insight update rate of 0.5 PPS. Different Insight tiers satisfy this requirement under different bandwidth conditions; for example, the High-Accuracy tier requires at least 11.68 Mbps, while lower-payload tiers can satisfy the same timeliness target under weaker bandwidth. Thus, if bandwidth falls below the High-Accuracy threshold, \texttt{AVERY} can still satisfy the timeliness requirement by switching to a lower-payload tier, such as Balanced, thereby trading some fidelity for continued timely updates.

\texttt{AVERY} does not solve an online numerical optimization problem. Instead, it enforces feasibility and selects a discrete operating point from a pre-profiled lookup table (LUT). When multiple feasible configurations exist, the controller resolves them according to mission priority: favoring higher-throughput configurations when timely updates are critical, or favoring higher-fidelity configurations when precise grounding is more important. In Section~\ref{sec:controller}, we operationalize this formulation through a lightweight deterministic controller that uses measured bandwidth together with the LUT to select a feasible runtime configuration.
\vspace{-2mm}

\section{\texttt{\papername{}} Architecture}
\label{sec:architecture} 

\begin{figure*}
\centering
\includegraphics[width=0.95\textwidth]{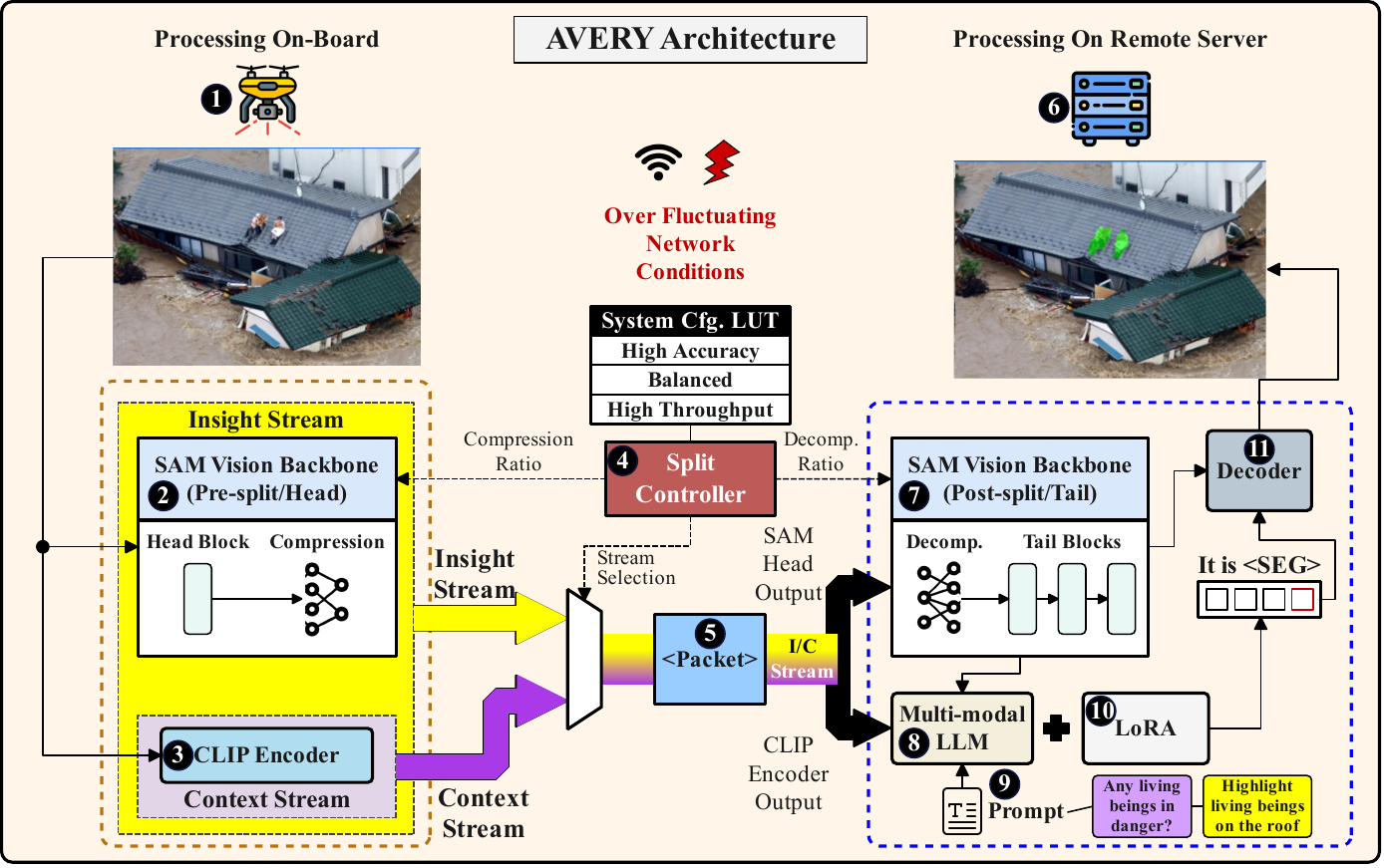}
\vspace{-2mm}
\caption{The \texttt{AVERY} architecture. The system partitions the VLM across onboard UAV processing and a remote server. Onboard, a dual-vision pipeline processes the captured image (\circlednum{1}) and produces features for two streams: a high-fidelity \textbf{Insight Stream} (bright yellow), derived from the SAM Vision Backbone (\circlednum{2}) and CLIP Encoder (\circlednum{3}), and a lightweight \textbf{Context Stream} (purple), derived from the CLIP Encoder alone (\circlednum{3}). The onboard Split Controller (\circlednum{4}) selects the admissible stream and, for the Insight Stream, applies a compression ratio chosen from a predefined LUT based on operator intent and network conditions. The selected representation is packetized (\circlednum{5}) and transmitted to the remote server (\circlednum{6}). On the server, the VLM combines SAM features (\circlednum{7}), CLIP features, and the operator prompt (\circlednum{9}) for reasoning in the Multi-Modal LLM (\circlednum{8}), while the Decoder (\circlednum{11}) generates the final segmentation mask when fine-grained grounding is required.} 
\vspace{2mm}
\label{fig:arch}
\vspace{-4ex}
\end{figure*}

We now present the \texttt{AVERY} architecture, illustrated in Figure~\ref{fig:arch}. 
The UAV captures images of a disaster zone \supc{1}, while the rescue operator issues language prompts remotely \supc{9}. 
The UAV and the remote operator communicate over a network with fluctuating conditions.
We take LISA-7B~\cite{lisa} as the base VLM, which combines a vision backbone with a multi-modal LLM to interpret natural language queries and generate pixel-level segmentation masks. \texttt{AVERY} deploys \textbf{split computing} to partition the LISA pipeline between the onboard UAV and a remote server. 

Unlike prior split computing works that primarily split a single model along its depth, splitting a VLM is more challenging because cross-modal reasoning introduces functional dependencies that are not naturally preserved by naive depth-based partitioning. More importantly, not all operator queries require the same semantic capability. Some queries require only coarse semantic awareness to support rapid triage, while others require pixel-level spatial grounding to support decisive action. \texttt{AVERY} therefore introduces a \textit{functional split} through a \textbf{dual-stream architecture}: a lightweight, high-frequency \textbf{Context Stream} (purple path in Figure~\ref{fig:arch}) for continuous situational awareness, and a heavyweight, lower-frequency \textbf{Insight Stream} (bright yellow path in Figure~\ref{fig:arch}) for high-fidelity spatial analysis. The orchestration of these streams is managed by an adaptive onboard controller (Figure~\ref{fig:arch}~\supc{4}), which uses operator intent to select the admissible stream and, for Insight operation, chooses the operating tier based on mission goal and runtime resource conditions.

At a high level, \texttt{AVERY} partitions the VLM across the UAV and a remote server as follows. On the UAV, the captured image is transformed into one of two transmitted representations: a lightweight \textbf{Context representation}, derived from the CLIP encoder alone, or a richer \textbf{Insight representation}, formed from both SAM and CLIP features. The Split Controller then determines which representation is admissible for the current query and, for the Insight path, which operating tier should be used before packetization and transmission. On the server, the received representation is fused with the operator prompt inside the multi-modal LLM, and when fine-grained grounding is needed, the decoder produces the final segmentation mask.

\begin{figure}[t]
\centering
\includegraphics[width=0.55\textwidth]{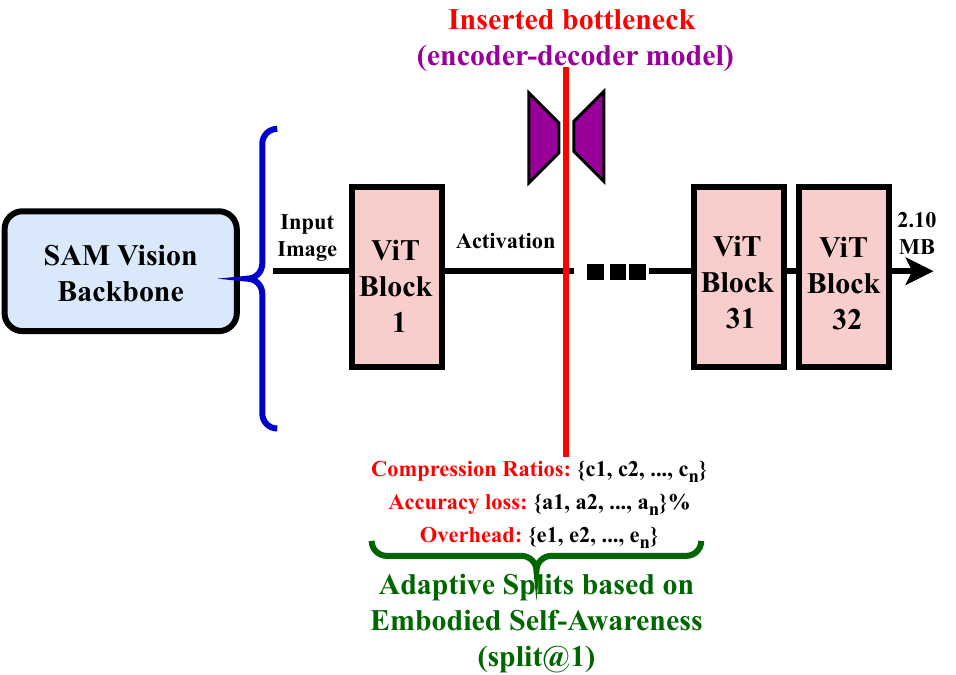}
\vspace{-3mm}
\caption{The \texttt{AVERY} split across model depth and compression mechanism. We insert a trainable bottleneck~\cite{matsubara2022bottlefit} (encoder--decoder pair) after the first ViT block (\texttt{split@1}) to compress the large SAM activation tensor (10.49 MB). Each pre-trained bottleneck model provides a different accuracy--throughput trade-off, forming the operational tiers used by the runtime controller.}
\vspace{-4mm}
\label{fig:split}
\end{figure}

\vspace{-2mm}
\subsection{Onboard Processing and the Split Controller} 

As shown in Figure~\ref{fig:arch}, the onboard pipeline transforms a captured high-resolution image~\supc{1} into representations for two distinct operating modes:
\begin{enumerate} 
\item \textbf{Context Stream:} Generated by a lightweight \textbf{CLIP Encoder}~\cite{clipx}~\supc{3}, this stream produces a compact, low-resolution semantic representation of the scene. It is computationally inexpensive and designed for frequent transmission, enabling rapid text-only reasoning and continuous situational updates. In effect, the Context Stream supports queries where the operator needs fast semantic awareness rather than precise localization.

\item \textbf{Insight Stream:} The high-fidelity path is built on the \textbf{Segment Anything Model (SAM)~\cite{kirillov2023segment} Vision Backbone}~\supc{2}, together with CLIP features from the same input image. We place the split point early, after the first vision transformer block (\texttt{split@1}), so that the UAV executes only a small prefix of the SAM backbone while the remaining visual processing is deferred to the server. The transmitted Insight packet therefore contains two components: compressed intermediate SAM activations, which preserve rich spatial structure for downstream mask generation, and CLIP-derived semantic features, which provide complementary scene-level context to the multi-modal LLM. This combination allows the server to reconstruct a richer visual-semantic state and produce grounded outputs when required.
\end{enumerate} 

The central onboard component is the adaptive \textbf{Split Controller}~\supc{4}. Its role is hierarchical. It first determines whether the operator's query requires coarse semantic response or fine-grained spatial grounding, and thereby selects the admissible stream. If the query can be satisfied semantically, the controller uses the lightweight Context Stream. If the query requires precise localization, the controller activates the Insight Stream. For Insight operation, it then consults the \textbf{System Configuration LUT} to choose the operating tier, which determines the bottleneck compression ratio and hence the size--quality operating point of the transmitted SAM activations, based on mission goal, network condition, and onboard resource budget. The selected packet is then transmitted~\supc{5} over the fluctuating network. Compression is performed by a trainable bottleneck~\cite{matsubara2022bottlefit}, as shown in Figure~\ref{fig:split}. 


This design is important because it separates two otherwise entangled decisions: \emph{what} semantic capability is required, and \emph{how} that capability should be delivered under current runtime constraints. \texttt{AVERY} therefore avoids both over-provisioning---e.g., unnecessarily sending expensive Insight packets for simple awareness queries---and under-provisioning, where lightweight updates would be insufficient for grounded tasks.

\subsection{Remote Server Processing: Reasoning and Segmentation} 

Once a packet is received by the remote server~\supc{6}, it is unpacked and processed by the tail of the VLM. The multi-modal LLM~\supc{8} forms the reasoning core, combining semantic context from the CLIP features with the operator's \textbf{Prompt}~\supc{9}. To improve domain specificity, we use \textbf{LoRA}~\cite{lora}~\supc{10} for parameter-efficient fine-tuning. The LLM generates a textual response that includes a special \texttt{<SEG>} token, which triggers the final decoding step. The \textbf{Decoder}~\supc{11} then uses the reconstructed high-fidelity SAM features---after bottleneck decoding---together with the LLM output to generate the final pixel-level segmentation mask.

The remote execution path therefore completes the portion of the model that is too expensive to execute continuously on the UAV. In the Context case, the server primarily performs lightweight semantic reasoning over CLIP-derived representations. In the Insight case, it performs richer multi-modal reasoning and spatial decoding over the combined SAM+CLIP representation. This asymmetric execution is what allows \texttt{AVERY} to support both timely awareness and detailed grounded analysis within one unified framework.

\subsection{Dual-Stream Operation in Practice} 

\texttt{AVERY}'s architecture supports a responsive workflow in which rapid triage can be followed by detailed spatial analysis. Here, operator intent determines whether the task can be served through lightweight semantic updates or requires full spatial grounding. For example, in the flood scenario shown in Figure~\ref{fig:arch}: 
\begin{itemize}
    \item The operator can use the high-frequency \textbf{Context Stream} to rapidly survey the scene and ask broad questions such as \texttt{``Are there any living beings on the rooftops?''} Based on frequent CLIP-only updates, the VLM can return a fast text response (e.g., \texttt{``Yes, two possible life signs detected on the grey roof''}), enabling efficient reconnaissance and filtering of low-priority regions.
    \item Once a high-priority target is identified, the operator can request detailed analysis: \texttt{``Highlight the living beings on that roof.''} The controller then activates the \textbf{Insight Stream} and selects the appropriate operating tier. A moment later, the server returns a precise segmentation mask. This enables the operator to distinguish, for instance, between a human survivor and an animal on the roof, supporting more informed resource-deployment decisions.
\end{itemize}

Viewed this way, the Context Stream is not simply a lower-cost approximation of Insight. Rather, it serves a distinct operational role: maintaining timely semantic awareness so that the operator can decide \emph{when} deeper grounded analysis is worth invoking. The Insight Stream, in turn, serves as the escalation path for queries whose value depends on precise spatial output. Together, the two streams allow \texttt{AVERY} to align VLM execution with the semantic demands of the task instead of treating all queries identically.

\subsection{Adaptive Control Logic for the Insight Stream}
\label{sec:controller}

We now present the adaptive control logic (Algorithm~\ref{alg:controller}) for \texttt{AVERY}'s onboard controller, which realizes the hierarchical runtime adaptation described in Section~\ref{sec:sysmodel}. The controller combines a pre-profiled system knowledge base with a lightweight deterministic policy to select the appropriate operating mode at runtime.
\begin{table}[tb!]
\centering
\caption{\texttt{AVERY} System Lookup Table (LUT)}
\vspace{-4mm}
\label{tab:lut}
\begin{tabular}{l c c c r}
\toprule
 & & \multicolumn{2}{c}{\textbf{Accuracy (Average IoU)}} & \\
\cmidrule(lr){3-4}
\textbf{Tier} & \textbf{\makecell{Compression \\ Ratio ($r$)}} &
\textbf{\makecell{Base/Original \\ Model}} &
\textbf{\makecell{Fine-tuned \\ Model}} &
\textbf{\makecell{Data \\ Size (MB)}} \\
\midrule
High Accuracy   & 0.25 & 84.42\% & 81.12\% & 2.92  \\
Balanced        & 0.10 & 82.89\% & 79.20\% & 1.35  \\
High Throughput & 0.05 & 80.67\% & 78.48\% & 0.83  \\
\bottomrule
\end{tabular}%
\vspace{-2mm}
\end{table}

\subsubsection{Knowledge Base (LUT)}
The controller relies on a static LUT (Table~\ref{tab:lut}) generated through offline profiling of the original and fine-tuned LISA models. For each Insight operating tier, the LUT stores: (i) compression ratio $r$, (ii) expected segmentation quality measured using Average IoU,  a heuristic localization quality proxy defined as the mean of gIoU and cIoU~\cite{lisa}, and (iii) compressed data size. Together, these entries define the operating points available to the runtime controller.

\subsubsection{Adaptive Policy}

Algorithm~\ref{alg:controller} implements \texttt{AVERY}'s runtime controller as a lightweight deterministic policy over a pre-profiled LUT. Its inputs are the current bandwidth $B_{\text{curr}}$, onboard compute-power budget $P_{\text{cfg}}$, mission goal $G_{\text{mission}}$, operator intent $I_t$, the intent-specific timeliness requirement $F_I$, and the system LUT $L_{\text{sys}}$, while its outputs are the selected configuration $C^*$ and its induced throughput $f^*$.\footnote{As the formal model includes the onboard compute-power budget as part of the runtime state, in the current prototype $P_{\text{cfg}}$ is instantiated through the selected Jetson Xavier operating mode and remains fixed during a deployment run. Unless otherwise stated, we use \texttt{MODE\_30W\_ALL}.}

The algorithm proceeds in four phases. In \textbf{Sense} (Line~10), the controller acquires the current bandwidth condition. In \textbf{Gate} (Lines~11--18), it determines whether the operator query corresponds to a Context-level or Insight-level intent. If the query only requires Context-level response, the controller selects the lightweight Context Stream, assigns the default Context configuration, and immediately returns the corresponding maximum Context throughput (Lines~14--17). This early return avoids unnecessary evaluation of Insight tiers for semantically lightweight queries.

If the query requires Insight-level processing, the controller enters \textbf{Evaluate} (Lines~19--28). It first initializes an empty feasible set $O_{\text{feas}}$ (Line~19), then iterates over all profiled Insight tiers in the LUT (Lines~20--25). For each tier $T_i$, it estimates the maximum achievable throughput $f_{i,\max}$ from the current bandwidth and the tier-specific compressed payload size (Line~21). A tier is inserted into the feasible set only if its achievable throughput satisfies the minimum update-timeliness requirement $F_I$ (Lines~22--23). If no Insight tier satisfies this constraint, the controller reports that no feasible Insight configuration exists under the current runtime condition (Lines~26--28). In the present prototype, this feasibility check is driven primarily by communication constraints, while $P_{\text{cfg}}$ captures the fixed onboard operating mode under which the controller executes.

Finally, in \textbf{Select} (Lines~29--35), the controller chooses among the feasible Insight tiers according to the mission goal. Under \texttt{PRIORITIZE\_ACCURACY}, it selects the highest-fidelity tier in $O_{\text{feas}}$ (Lines~29--30). Under \texttt{PRIORITIZE\_THROUGHPUT}, it selects the highest-throughput feasible tier (Lines~31--32). The controller then retrieves the throughput induced by the selected configuration (Line~34) and returns the final pair $(C^*, f^*)$ (Line~35). In this way, \texttt{AVERY} enforces semantic admissibility first, filters configurations by timeliness feasibility second, and only then applies mission-aware preference over the remaining operating points.
\begin{algorithm}[tbh]
\small
\caption{\texttt{AVERY} Onboard Controller Logic}
\label{alg:controller}
\begin{algorithmic}[1]
\Procedure{SelectConfiguration}{$B_{\text{curr}},\, P_{\text{cfg}},\, G_{\text{mission}},\, I_t,\, F_I,\, L_{\text{sys}}$}
    \State \textbf{Input:}
        $B_{\text{curr}}$: Current bandwidth (Mbps). \\
        $P_{\text{cfg}}$: Onboard compute-power budget / operating mode. \\
        $G_{\text{mission}}$: Mission goal from \{``Accuracy'', ``Throughput''\}. \\
        $I_t$: Operator intent (Context-level vs. Insight-level). \\
        $F_I$: Minimum update-timeliness requirement for intent $I_t$. \\
        $L_{\text{sys}}$: System LUT (Table~\ref{tab:lut}).
    \State \textbf{Output:}
        $C^*$: Selected configuration. \\
        $f^*$: Induced throughput (PPS).

    \Statex
    \Comment{\textbf{Stage 1: Sense}}
    \State Acquire current bandwidth $B_{\text{curr}}$.

    \Statex
    \Comment{\textbf{Stage 2: Gate}}
    \If{$I_t$ is an Insight-level intent}
        \State $S_t \gets$ \textsc{InsightStream}
    \Else
        \State $S_t \gets$ \textsc{ContextStream}
        \State $C^* \gets$ \textsc{ContextConfiguration}
        \State $f^* \gets$ \textsc{MaxContextThroughput}
        \State \Return $(C^*, f^*)$
    \EndIf

    \Statex
    \Comment{\textbf{Stage 3: Evaluate Feasible Insight Tiers}}
    \State Let $O_{\text{feas}}$ be an empty set of feasible operating points.
    \For{each tier $T_i$ in $L_{\text{sys}}$}
        \State Estimate $f_{i,\max} \gets (B_{\text{curr}} / 8) / T_i.\text{data\_size}$
        \If{$f_{i,\max} \ge F_I$}
            \State Add $(T_i, f_{i,\max})$ to $O_{\text{feas}}$
        \EndIf
    \EndFor

    \If{$O_{\text{feas}}$ is empty}
        \State \Return \textsc{NoFeasibleInsightTier}
    \EndIf

    \Statex
    \Comment{\textbf{Stage 4: Select Tier by Mission Goal}}
    \If{$G_{\text{mission}}$ is \texttt{``PRIORITIZE\_ACCURACY''}}
        \State $C^* \gets$ highest-fidelity tier in $O_{\text{feas}}$
    \ElsIf{$G_{\text{mission}}$ is \texttt{``PRIORITIZE\_THROUGHPUT''}}
        \State $C^* \gets$ highest-throughput tier in $O_{\text{feas}}$
    \EndIf

    \State $f^* \gets$ throughput corresponding to $C^*$ in $O_{\text{feas}}$
    \State \Return $(C^*, f^*)$
\EndProcedure
\end{algorithmic}
\end{algorithm}

%% file: tex/05_Experiments.tex
\vspace{-3ex}
\section{Experiments and Evaluation}
\label{sec:experiments}
We now present a comprehensive evaluation of \texttt{AVERY}. Our evaluation is structured in two parts. First, we present a static characterization of the VLM to motivate our core design choices, such as the fixed split point and the dual-stream architecture. Second, we evaluate the full adaptive \texttt{AVERY} system (Insight Stream) against static baselines under dynamic network conditions to quantify the benefits of its self-aware control logic.

\subsection{Experimental Setup}
\subsubsection{Hardware and Software}
Our experimental testbed consists of an NVIDIA Jetson AGX Xavier (32 GB) as the onboard UAV computer, a platform widely used for autonomous edge agents~\cite{bertozzi, access_av}, configured to operate in its \texttt{MODE\_30W\_ALL} power mode. The remote server is equipped with an NVIDIA RTX 6000 Ada Generation GPU. For the VLM, we use LISA-7B~\cite{lisa} with FP16 precision~\cite{fp}. To ensure a stable and representative workload for our system-level evaluation, we use the pre-trained Original LISA model. To model disaster scenarios, we also use the fine-tuned model that uses the \texttt{Flood-ReasonSeg} dataset, as detailed below.
\begin{figure}[t]
    \centering
    \includegraphics[width=\linewidth]{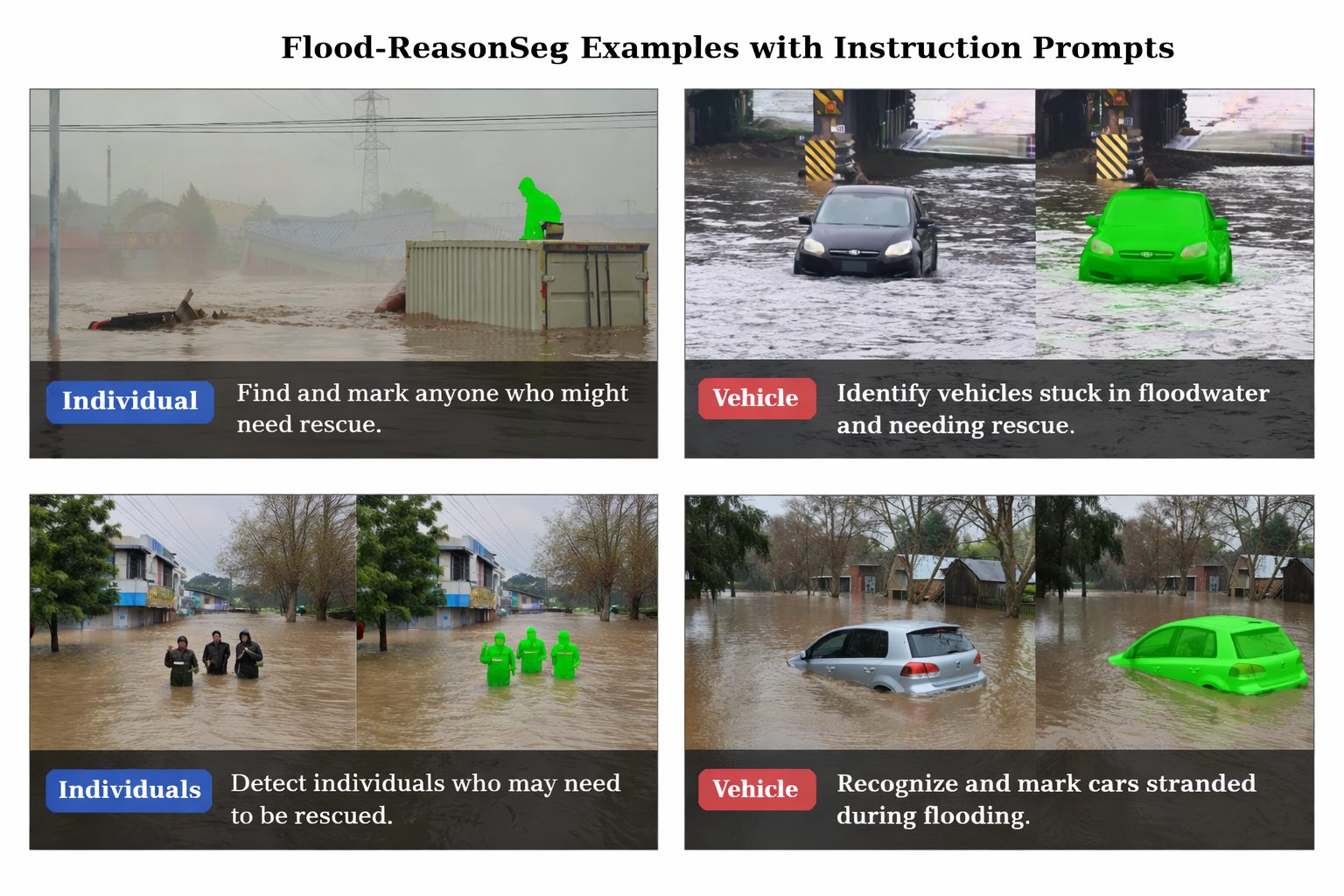}
    \caption{\textbf{Examples from \texttt{Flood-ReasonSeg}.} Each flood-scene image is annotated in \texttt{ReasonSeg} format using a natural-language instruction and a segmentation mask. The dataset targets two flood-response classes: stranded individuals requiring rescue and vehicles stranded by floodwater.}
    \label{fig:flood_reasonseg_examples}
\end{figure}

\subsubsection{\texttt{Flood-ReasonSeg} Dataset}

To capture flood-response needs, we curated \texttt{Flood-ReasonSeg}, a dataset targeting two classes: individuals requiring rescue and vehicles stranded by water. About 100 flood images were collected and split into 70\% training / 30\% validation. Photometric data augmentation~\cite{photo, photo2} expanded training to $\sim$300 samples (enough to fine-tune~\cite{lisa}). Each image was annotated in \texttt{ReasonSeg} format~\cite{lisa} with natural-language instructions and segmentation masks (Figure~\ref{fig:flood_reasonseg_examples}). We fine-tuned LISA using LoRA, producing a flood-specialized VLM.

\vspace{-2ex}
\subsection{Static System Characterization} As we are the first to investigate split computing for VLMs in this context, we first performed a detailed profiling study to understand the system's performance characteristics and justify our final architectural design.

\subsubsection{Model Depth Splitting} 
\begin{figure}[t]
\centering
\includegraphics[width=0.75\textwidth]{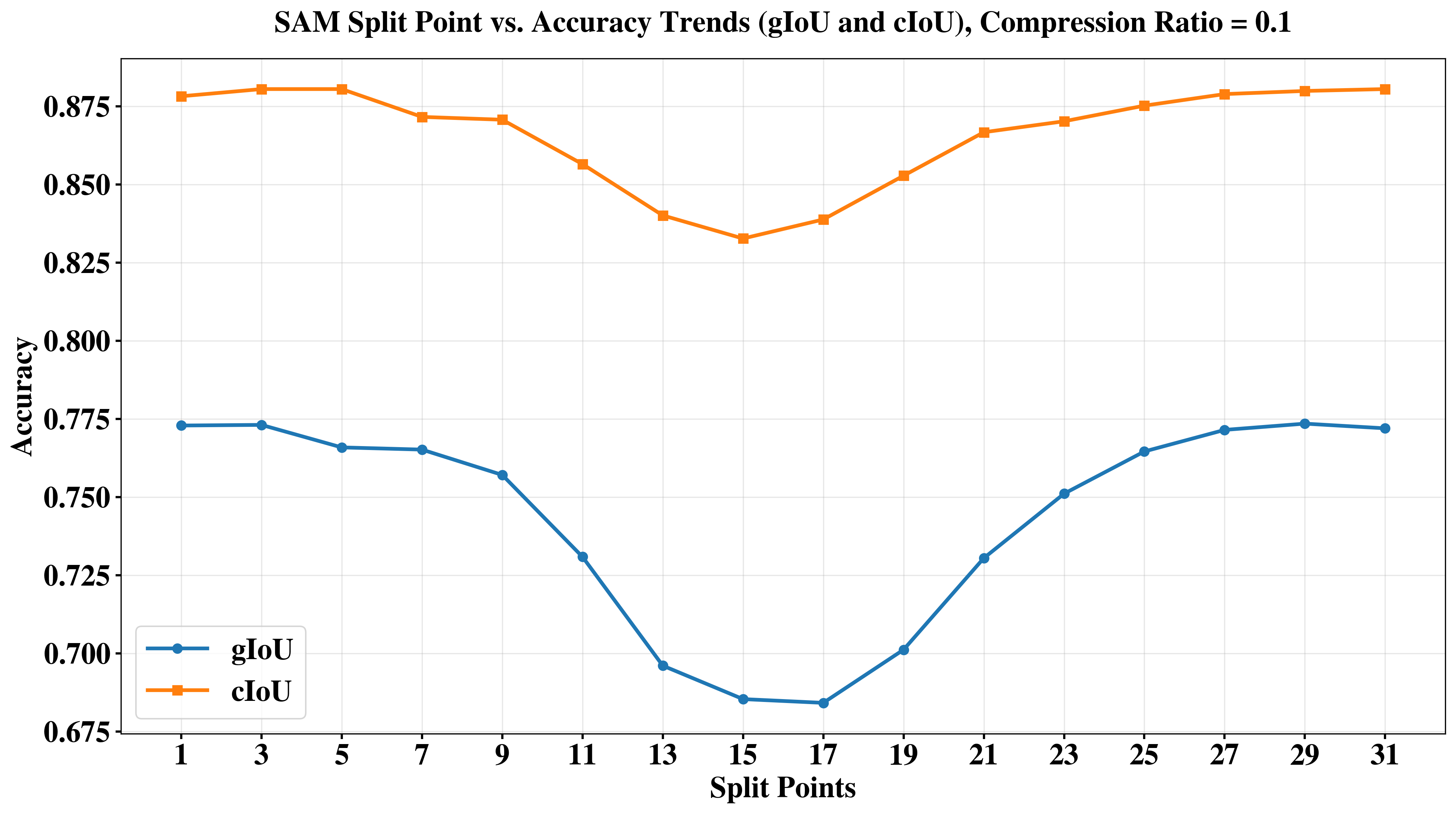}
\vspace{-2mm}
\caption{SAM split-point accuracy trends at compression ratio $r=0.1$. Although some later split points maintain competitive gIoU and cIoU, deeper splits push more of the SAM backbone onto the UAV and therefore incur substantially higher on-device energy cost. \texttt{AVERY} therefore favors an early split point that provides a better overall accuracy--energy trade-off for bottlenecked~\cite{matsubara2022bottlefit} execution.}
\vspace{2mm}
\label{fig:trend}
\end{figure}

\begin{figure}[t]
\centering
\includegraphics[width=0.75\textwidth]{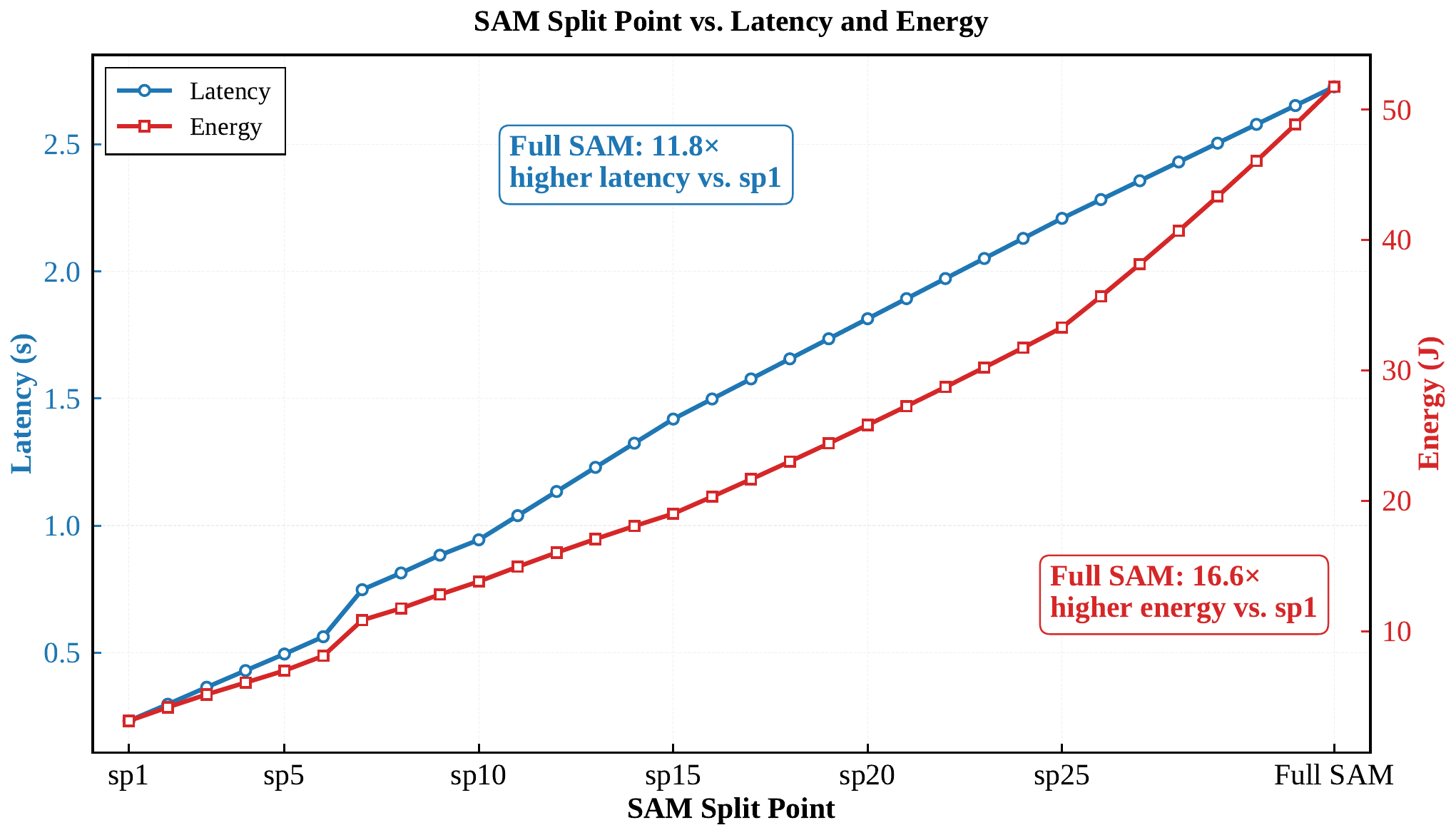}
\vspace{-2mm}
\caption{\textbf{Latency and energy per image across SAM split points on Jetson AGX Xavier (MODE\_30W\_ALL).} Here, \texttt{spk} denotes a split after the $k$-th ($k = [1,31]$, $k \in N$) ViT block of SAM, while \textit{Full SAM onboard} represents full local execution of the SAM pipeline on the device. Compared to \texttt{sp1}, \textit{Full SAM onboard} increases latency by $11.8\times$ and energy per frame by $16.6\times$.}
\vspace{2mm}
\label{fig:laten}
\end{figure}

We quantified the benefits of splitting LISA's SAM~\cite{kirillov2023segment} vision backbone across its depth. The accuracy trends across split depths are shown in Figure~\ref{fig:trend}, while the corresponding latency and energy trends are shown together in Figure~\ref{fig:laten}. From this analysis, we make \textbf{four} major observations. \textbf{First}, compared to running the full SAM backbone on the Jetson, our \texttt{split@1} strategy, which places the split after the first ViT block, reduced energy per frame by \textbf{93.98\%}, directly extending UAV endurance. \textbf{Second}, \texttt{split@1} also delivered \textbf{11.2\% higher accuracy} than raw image compression\footnote{The initial ViT block acts as a powerful feature extractor~\cite{feature}; our bottleneck therefore compresses a representation already distilled for task-salient information, which is more efficient than a bottleneck on raw pixels that must learn to both extract features and compress them simultaneously.}. \textbf{Third}, deeper splits did not provide a compelling accuracy-efficiency tradeoff, as they initially degraded accuracy while sharply increasing edge cost. For example, moving from \textbf{ViT-1} to \textbf{ViT-11} reduced average accuracy from \textbf{0.8256} to \textbf{0.7937}, corresponding to a \textbf{3.86\%} relative drop, while increasing on-device latency by \textbf{307.29\%} \textbf{(0.2318 s $\rightarrow$ 0.9441 s)} and on-device energy consumption by \textbf{342.63\%} \textbf{(3.12 J $\rightarrow$ 13.81 J)}. Accuracy then dipped further at intermediate depths, reaching \textbf{0.7615} at \textbf{ViT-17}. \textbf{Fourth}, although accuracy gradually recovered in later layers, this recovery was negligible from a systems perspective. At \textbf{ViT-29}, the average accuracy reached \textbf{0.8267}, which is only a \textbf{0.14\%} improvement over \textbf{ViT-1}, yet it required \textbf{43.34 J} per frame and \textbf{2.5044 s} of on-device latency, corresponding to \textbf{1290.23\%} higher energy consumption and \textbf{980.41\%} higher latency than \texttt{split@1}. Running the full SAM backbone on the Jetson was even more expensive at \textbf{51.75 J} and \textbf{2.7262 s} per frame. We therefore fixed the split at \textbf{ViT-1} \textbf{(\texttt{split@1})}, since it captures the best overall tradeoff across accuracy, latency, and energy, and avoids paying a large edge-side cost for negligible late-stage accuracy gains.

 \subsubsection{Dual-Stream Architecture} 
 \label{sec:context}
\texttt{split@1} produces a high-fidelity but low-frequency Insight Stream due to compute and transmission overhead. To preserve situational awareness, we added a lightweight Context Stream. Profiling showed that on-device processing for the CLIP-based Context Stream is \textbf{6.4$\times$ faster} than the Insight Stream on average, enabling continuous feedback while Insight provides periodic intelligence updates.

\vspace{-2ex}
\subsection{Adaptive System Evaluation (Insight Stream)}
\begin{figure*}
\centering
\includegraphics[width=0.9\textwidth]{images/merged_grid.pdf}
\vspace{-2mm}
\caption{Runtime results of \texttt{AVERY}'s Insight Stream in ``Prioritize Accuracy'' mode (20 minutes). (a) Bandwidth variation over time; (b) Runtime tier switching between High Accuracy and Balanced modes; (c) Accuracy comparison for both original and fine-tuned models; (d) Throughput comparison among different tiers and \texttt{AVERY}.}
\label{fig:results}
\vspace{2mm}
\vspace{-4ex}
\end{figure*}

\subsubsection{Evaluation Methodology}
We conduct a 20-minute experiment representative of typical UAV flight duration~\cite{skydio} under simulated dynamic network conditions. To rigorously test adaptability, our scripted trace emulates a disaster environment with stable periods, high volatility, and sustained drops, all within an 8–20 Mbps range—typical of uplink capacity in developing regions—which we use as a \textit{proxy} for degraded 5G connectivity in disaster zones~\cite{africa_bw, bw2}. Both Original and flood-related datasets are streamed in round-robin fashion. 
We compare \texttt{AVERY} against three static baselines fixed to one tier: \textbf{High-Accuracy}, \textbf{Balanced}, and \textbf{High-Throughput}\footnote{Though we show 3 tiers for \texttt{AVERY} with rule-based control, future work will involve more advanced control policies with higher granularity.}. Metrics include average IoU, a heuristic localization-quality proxy defined as the mean of gIoU and cIoU~\cite{lisa}, and throughput (Packets Per Second, PPS).

 \subsubsection{Results} Figure~\ref{fig:results} summarizes our 20-minute dynamic evaluation. We make 3 major observations. (1) In ``Prioritize Accuracy'' mode, \texttt{AVERY} adapts at runtime by identifying when the High-Accuracy tier becomes infeasible and switching between the High-Accuracy and Balanced tiers (Figure~\ref{fig:results}(b)).  
(2) This adaptability yields more stable throughput than the brittle High-Accuracy baseline, which collapses under low bandwidth (Figure~\ref{fig:results}(d)).  
(3) Accuracy remains robust: \texttt{AVERY}’s average IoU is within 0.75\% of the High-Accuracy baseline, while consistently outperforming Balanced and High-Throughput tiers (Figure~\ref{fig:results}(c)).

\begin{figure}
\centering
\includegraphics[width=0.65\textwidth]{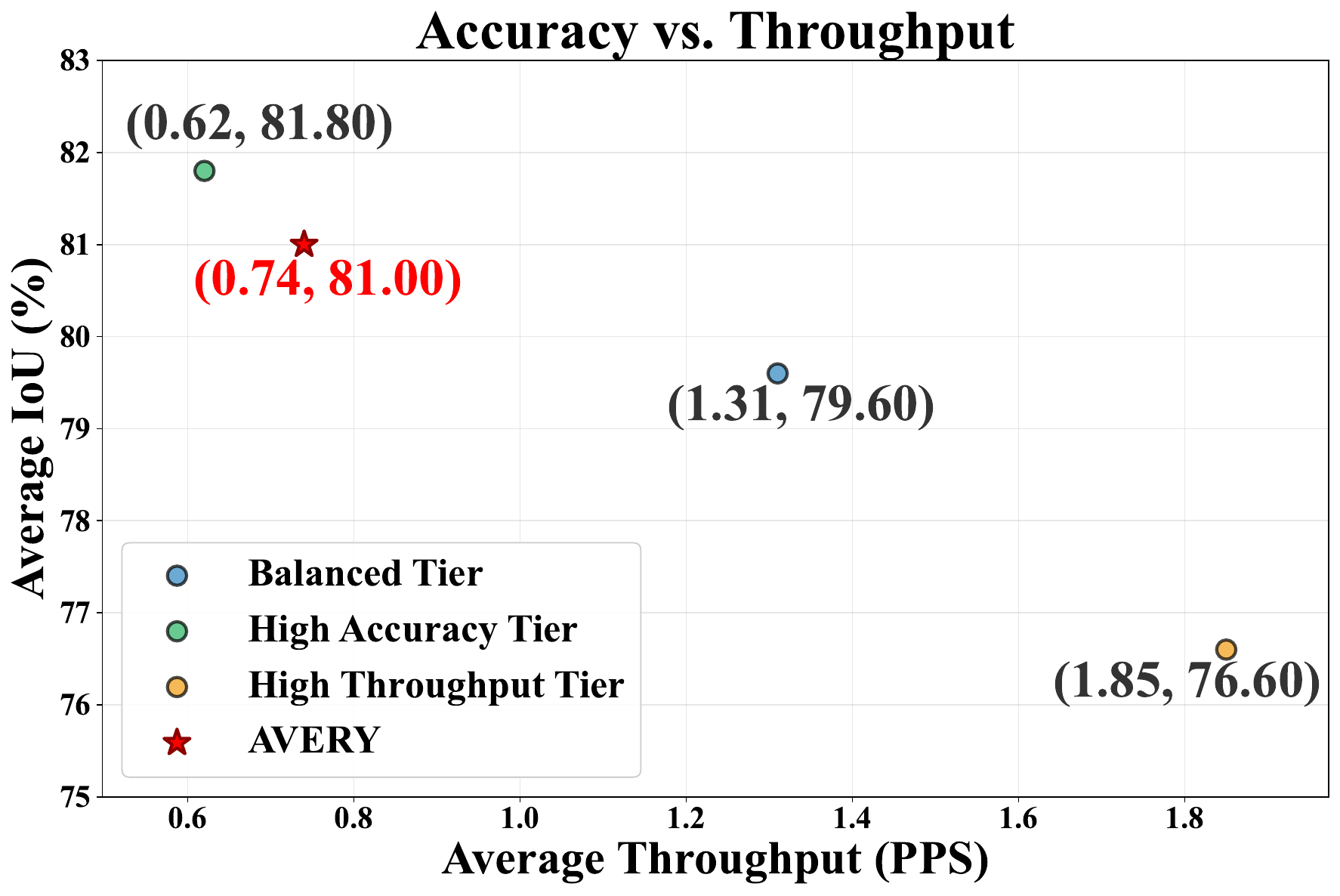}
\vspace{-2mm}
\caption{{Trade-off Analysis. Average Accuracy vs. Average Throughput for different operational tiers (Insight Stream, AVERY is in "Prioritize Accuracy" Mode. Results are for the Original model). }}
\vspace{2mm}
\label{fig:dse}
\vspace{-4ex}
\end{figure}

Figure~\ref{fig:dse} highlights the core trade-off. By making small, controlled accuracy sacrifices, \texttt{AVERY} sustains a stable throughput of 0.74 PPS, directly translating into an energy-efficient operating point that extends UAV endurance—a blended profile unattainable by any static configuration. {When throughput is prioritized, the Insight Stream reaches 1.85 PPS, while the much faster Context Stream (as previously quantified in Section~\ref{sec:context}) easily maintains real-time situational awareness, giving \texttt{AVERY} a balanced and robust operating envelope.}\\

%% file: tex/07_conclusion.tex
\section{Conclusion and Future Work}
\label{sec:conclusion}
We presented \texttt{AVERY}, to the best of our knowledge, the first intent-driven adaptive VLM split computing framework for disaster-response UAVs. The central idea in \texttt{AVERY} is that edge--cloud execution should not be determined by resource conditions alone: different operator queries demand different semantic products, and therefore different execution pathways. By combining early edge--cloud partitioning, learned bottleneck compression, a dual-stream architecture, and lightweight runtime control, \texttt{AVERY} aligns VLM execution with both semantic need and runtime feasibility.

Our results show that this design yields a practical operating point for embodied deployment. Across dynamic bandwidth variation, \texttt{AVERY} sustains higher throughput than fixed high-accuracy operation while remaining within $\sim$0.75\% of the static High-Accuracy baseline, and reduces on-device energy by 93.98\% relative to full edge execution of the Insight path. More broadly, the system demonstrates that semantically distinct execution modes are useful in practice: the lightweight Context Stream supports timely situational triage, while the richer Insight Stream is invoked only when precise spatial grounding is required. In this sense, \texttt{AVERY} shows that VLM deployment for embodied agents should not be treated as a monolithic offloading problem, but as an intent-conditioned adaptive systems problem.

At the same time, several important directions remain open. Our current evaluation emphasizes adaptability, throughput, and energy, but fuller mission-level measures of operational effectiveness remain necessary. The present prototype is instantiated on LISA~\cite{lisa} and evaluated under a fixed onboard operating mode, so extending \texttt{AVERY} to more tightly fused VLM architectures~\cite{t1,t2,vlm2}, richer onboard resource envelopes, and broader embodied settings remains an important next step. Likewise, while the current controller is intentionally lightweight and interpretable, more advanced policies~\cite{mpc,b1,rl2} may better exploit richer context signals. Complementary techniques such as pruning~\cite{prr,tinoosh} and quantization~\cite{ispa,tsar} may further reduce both onboard and transmission cost, and extending the framework to multi-UAV coordination~\cite{multi,cord,sari} would help test whether intent-driven semantic adaptation remains beneficial at larger system scale. Overall, \texttt{AVERY} provides a first step toward making grounded VLM perception practical for resource-constrained embodied agents operating in adverse environments, and suggests a broader systems direction in which semantic intent, communication conditions, and onboard resource limits are treated jointly in the design of adaptive physical intelligence.

%% file: tex/ack.tex
\section{Acknowledgements}
We would like to thank Nalini Venkatasubramanian, Wenjun Huang, Kapil Agrawal, and Arnab Sarkar for their valuable inputs that helped this research. Part of the experimental evaluation was carried out on the Chameleon testbed~\cite{chameleon}, supported by the National Science Foundation.